\documentclass{PoS}

\usepackage{siunitx}

\title{Mini-EUSO engineering model: \\tests in open-sky condition}

\ShortTitle{Mini-EUSO engineering model: tests in open-sky condition}

\author{\small\speaker{F.~Bisconti}$^1$, D.~Barghini$^2$, M.~Battisti$^{1,3}$, A.~Belov$^{4}$, M.E.~Bertaina$^{1,3}$, S.~Blin-Bondil$^{5}$, F.~Cafagna$^{6}$, G.~Cambi\`e$^{7,8}$, F.~Capel$^9$, M.~Casolino$^{7,8,10}$, A.~Cellino$^{1,2}$, I. Churilo$^{11}$, G.~Cotto$^{1,3}$, A.~Djakonow$^{12}$, T.~Ebisuzaki$^{10}$, F.~Fausti$^{1,3}$, F.~Fenu$^{1,3}$, C.~Fornaro$^{13}$, A.~Franceschi$^{14}$,	C.~Fuglesang$^{9}$, D.~Gardiol$^2$, P.~Gorodetzky$^{15}$, F.~Kajino$^{16}$, P.~Klimov$^{4}$, L.~Marcelli$^{7}$,  W.~Marsza{\l}$^{12}$, M.~Mignone$^{1}$, H. Miyamoto$^{1,3}$,  A.~Murashov$^{5}$, T.~Napolitano$^{14}$, G.~Osteria$^{17}$, M.~Panasyuk$^{4}$, E.~Parizot$^{15}$, A.~Poroshin$^{4}$, P.~Picozza$^{7,8}$, L.W.~Piotrowski$^{10}$, Z.~Plebaniak$^{12}$, G.~Pr\'ev\^ot$^{15}$, M.~Przybylak$^{12}$, E.~Reali$^{8}$, M.~Ricci$^{14}$, N.~Sakaki$^{10}$, K.~Shinozaki$^{1,3}$, G.~Suino$^{3}$, J.~Szabelski$^{12}$, Y.~Takizawa$^{10}$, M.~Tra\"{i}che$^{18}$, and S. Turriziani$^{10}$ \newline for the JEM-EUSO Collaboration\footnote{for collaboration list see PoS(ICRC2019)1177}\\
$^1$INFN Turin, Italy; $^2$INAF-OATo Turin, Italy; $^3$University of Turin, Italy; $^4$SINP, Lomonosov Moscow State University, Moscow, Russia; $^5$Omega, Ecole Polytechnique, CNRS/IN2P3, Palaiseau, France, $^6$INFN Bari, Italy; $^7$INFN Roma Tor Vergata, Italy; $^8$University of Roma Tor Vergata, Italy; $^9$KTH Royal Institute of Technology, Stockholm, Sweden; $^{10}$RIKEN, Wako, Japan; $^{11}$Russian Space Corporation Energia, Moscow, Russia; $^{12}$National Centre for Nuclear Research, Lodz, Poland; $^{13}$UTIU Rome, Italy; $^{14}$INFN Laboratori Nazionali di Frascati, Italy; $^{15}$APC, Univ Paris Diderot, CNRS/IN2P3, CEA/Irfu, Obs de Paris, Sorbonne Paris Cit\'e, France; $^{16}$Konan University, Kobe, Japan; $^{17}$INFN Naples, Italy; $^{18}$CDTA Algiers, Algeria \\
E-mail: \email{francesca.bisconti@to.infn.it}}

\abstract{Mini-EUSO is a UV telescope that will look downwards to the Earth's atmosphere onboard the International Space Station. With the design of the ultra-high energy cosmic ray fluorescence detectors belonging to the JEM-EUSO program, it will make the first UV map of the Earth by observing atmospheric phenomena such as transient luminous events, sprites and lightning, as well as meteors and bioluminescence from earth. Diffused light from laser shots from the ground, which mimic the fluorescence light emitted by Nitrogen molecules when extensive air showers pass through the atmosphere, can be used to verify the capability of this kind of detector to observe ultra-high energy cosmic rays. To validate the electronics and the trigger algorithms developed for Mini-EUSO, a scaled down version of the telescope with 1:9 of the original focal surface and a lens of 2.5 cm diameter has been built. Tests of the Mini-EUSO engineering model have been made in laboratory and in open sky condition. In this paper, we report results of observations of the night sky, which include the detection of stars, meteors, a planet and a rocket body reflecting the sunlight. Interesting results of the observation of city lights are also reported.}

\FullConference{36th International Cosmic Ray Conference -ICRC2019-\\
		July 24th - August 1st, 2019\\
		Madison, WI, U.S.A.}

\begin{document}
\setcounter{page}{2}	

\section{Introduction}
Mini-EUSO \cite{bib:mini-euso,bib:mini-euso-icrc2019} is the first experiment within the JEM-EUSO program \cite{bib:jem-euso} to be operated in space on-board the International Space Station (ISS), which orbits around the Earth at an altitude of about 400~km. All the experiments of the JEM-EUSO program consist of UV detectors with high temporal and spatial resolution, designed for the detection of Ultra-High Energy Cosmic Rays (UHECRs) at nighttime. Mini-EUSO will be placed at a nadir-facing window transparent to the UV light, in the Russian Zvezda module. The main objective of the mission is to create the first UV map of the Earth from space, which is useful for future large-scale and space-based detectors for the observation of UHECRs. It will also provide data to study bioluminescence phenomena in the sea, atmospheric events such as transient luminous events (TLEs), meteors, as well as to search for strange quark matter.

A Mini-EUSO \textit{engineering model} (Mini-EUSO~\textit{EM}) was developed to test the design of Mini-EUSO, in particular the electronics and the trigger algorithm developed for the detection of different time-scale events, taking into account the multidisciplinary nature of the mission. It is a scaled version of Mini-EUSO, with 1:9 of the focal surface and a lens of diameter 2.5~cm instead of the optical system composed of two Fresnel lenses of diameter 25~cm. Instead, the electronics and the software mounted in it were the same as those foreseen for Mini-EUSO \textit{flight model}. Three nights of data-taking campaigns in open-sky condition at the Astronomical Observatory of Turin (Italy) and on the roof of the Department of Phyiscs of the University of Turin provided a large variety of data that have been analyzed and the results are reported in this proceedings. 

The Department of Physics hosts the TurLab facility \cite{bib:turlab_nim}, with a rotating tank used mainly for fluid dynamics studies, but that can be also used to emulate the Earth rotation by fixing the telescope on the roof to observe down towards the tank, where light reflected by different materials, such as bricks and glass, or those by LEDs, represent different background sources from Earth. Results of these studies are reported in \cite{bib:turlab}.

\section{The Mini-EUSO~\textit{engineering model}}
The Mini-EUSO~\textit{EM} consists of a lens of diameter 2.5~cm which focuses the light on a focal surface corresponding to the central Elementary Cell (EC) of the Mini-EUSO focal surface. One EC is made of $2\times2$ Multi-Anode Photo-Multiplier Tubes (MAPMTs) \cite{bib:mapmt}, each with $8\times8$~pixels of side 2.88~mm. The complete focal surface of Mini-EUSO, as well as the other experiments of the JEM-EUSO program, consist of an array of $3\times3$~ECs together with the electronics boards, which compose the Photo-Detector Module (PDM). A UV transmitting band pass filter in the range $290-430$~nm is glued on top of each MAPMT. 
\begin{figure}[h]
	\centering	
	\includegraphics[height=5.cm]{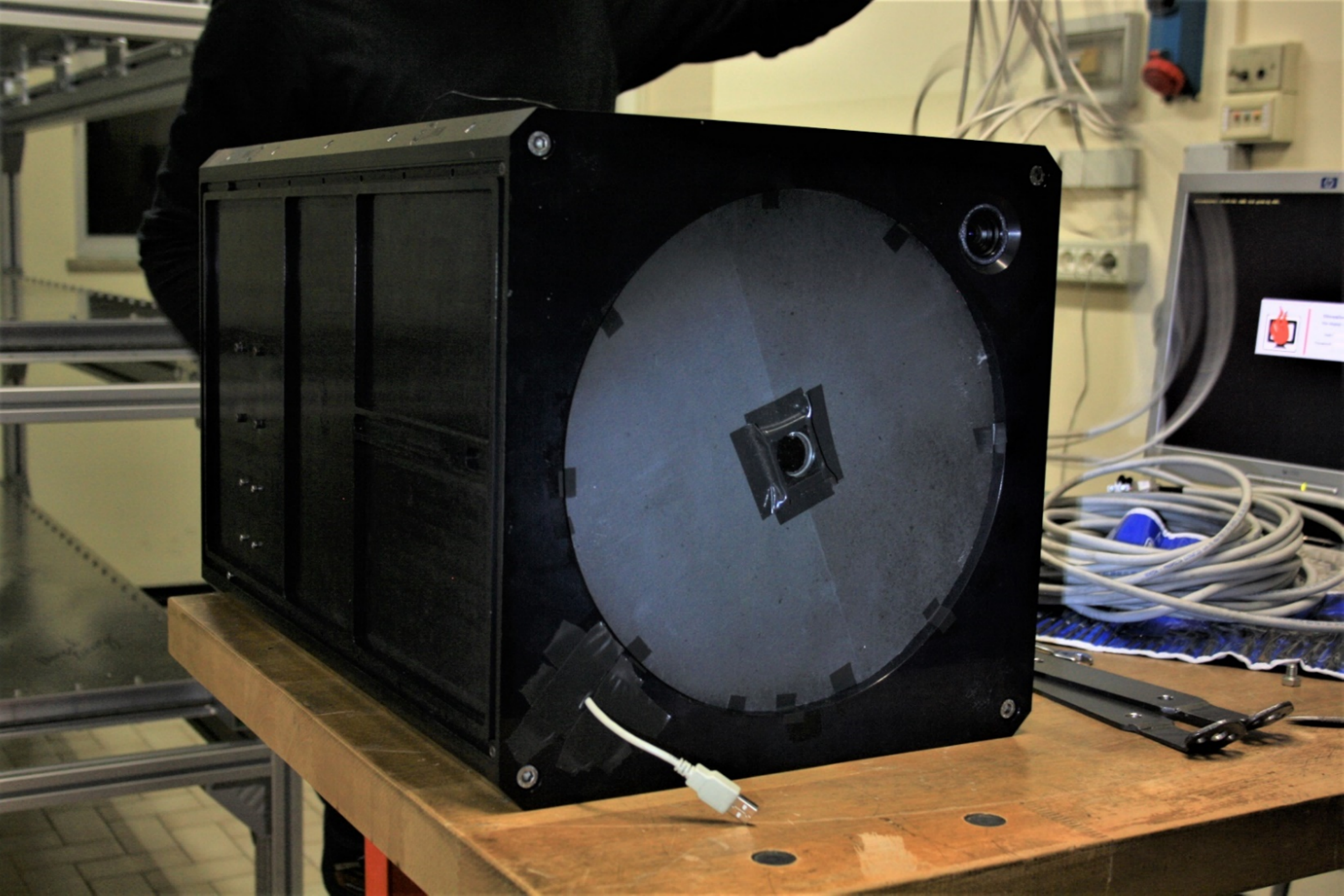}\hspace{2mm}
	\includegraphics[height=5.cm]{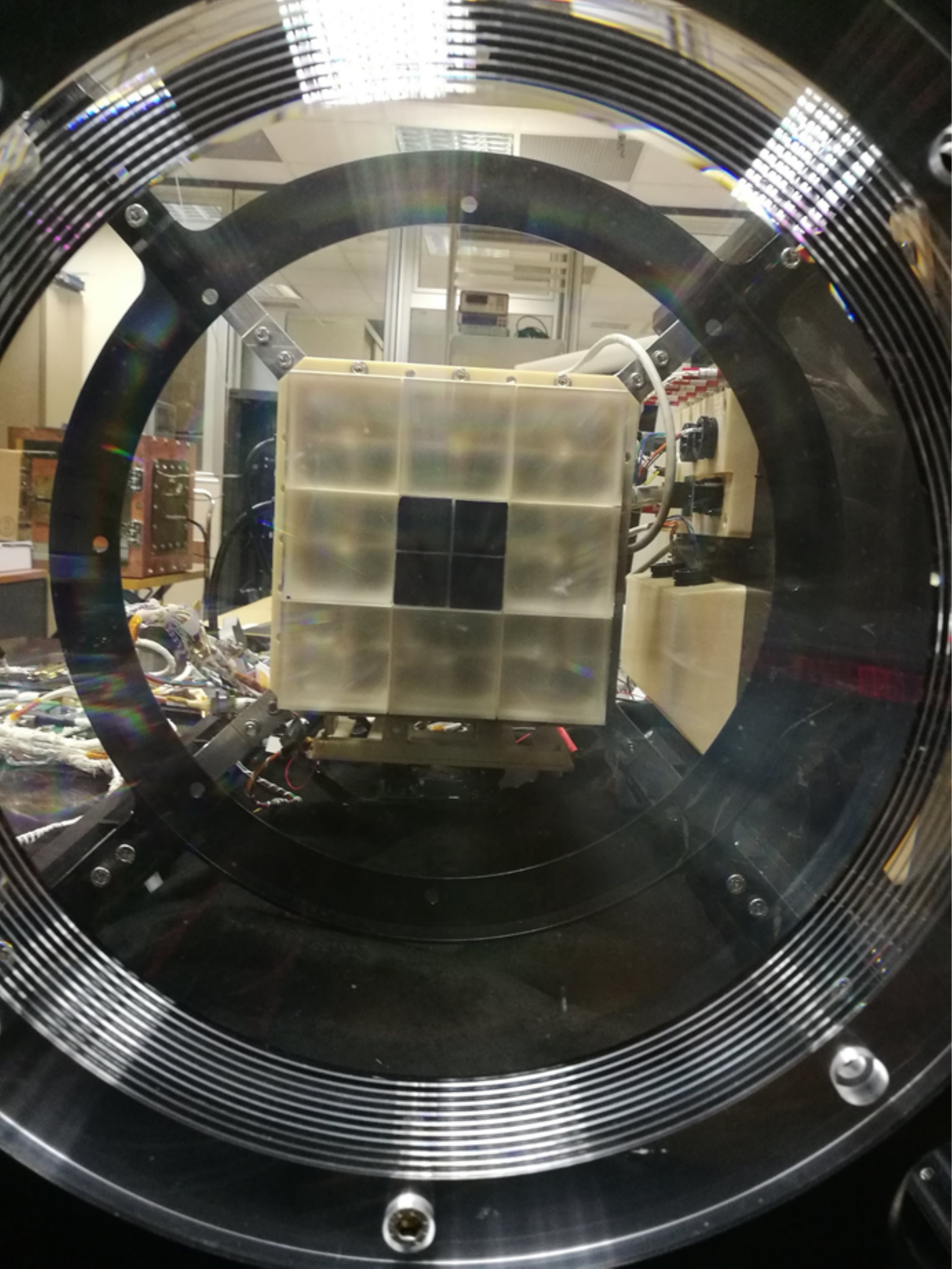}
	\caption{\textit{Left:} Mini-EUSO~\textit{EM} exterior sight, with the lens of diameter 2.5~cm; \textit{Right:} the internal focal surface made of $2\times2$~MAPMTs.} \label{fig:mini-euso-em}
\end{figure}\\
The left panel of Figure~\ref{fig:mini-euso-em} shows the exterior of Mini-EUSO~\textit{EM}, including the lens at the center of the front side. The right panel shows the focal surface, where there is the central EC of the PDM.

Data are sampled every Gate Time Units (GTUs), $2.5\,\mu\mbox{s}$ long. The GTU corresponds to the time in which the UHECR pass through the distance seen by a pixel of a MAPMT placed in space at 400~km altitude, and represents the time resolution of the detectors within the JEM-EUSO program. The readout is performed by one 64-channel SPACIROC3 ASIC chip \cite{bib:spaciroc3} per MAPMT. Data are saved with three time resolutions to cover the different timescales typical of the various phenomena Mini-EUSO aims to observe from space: 1 GTU (2.5~$\mu$s), $128\times 2.5\,\mu$s (320~$\mu$s), and $128\times 320\,\mu$s (40.96~ms) \cite{bib:zynq}.
%
\section{Sky observations}
Tests on the Mini-EUSO \textit{EM} were performed in March 2018 in Turin, at the Astronomical Observatory and at the Department of Phyiscs. There were a few advantages performing tests in these locations. 
The Astronomical Observatory of Turin hosts one camera of the PRISMA network \cite{bib:prisma} devoted to the observation of fireballs in the sky with the goal of recovering the remaining meteorites on ground and analyze their composition. The PRISMA cameras operates with a high frequency of 30~frames/s to follow the fireball development properly. Images integrated over 5~s are saved every 10~minutes, in order to see stars and calculate the position of the fireballs in the sky using star positions. 
In the nights of March 12-13, 2018, data acquisitions of the sky with Mini-EUSO~\textit{EM} were made from the Astronomical Observatory site. A comparison of the results of the simultaneous stars observations using the PRISMA camera and the Mini-EUSO~\textit{EM} was made to understand the relative performance of the two detectors, and also in view of possible future observations of meteors with both the PRISMA cameras and Mini-EUSO from the ISS. In addition, observations of faint meteors and a rocket body in orbit were performed. However, they were not triggered by the PRISMA camera as such slow events were rejected by the trigger algorithm. Moreover, observation of artificial light from nearby cities were made, too. In the night of March 14, 2018, observations of the open sky were performed on the roof of the Department of Physics, observing lighted signs and flashes from skyscrapers of the city and an airplane.
In this proceedings, the observation and relative analysis of several light sources are discussed, split in natural and artificial sources.

\subsection{Natural sources}

\subsubsection{Stars and planet}
Several stars and the Jupiter planet have been identified in the data taken with Mini-EUSO~\textit{EM} from the Astronomical Observatory by pointing at different zenith angles. From the relative position of three stars of the Hercules constellation visible at the same time in the data frames, the total field of view of the telescope resulted to be about $\pm$\ang{5}, or $\sim$\ang{0.6}/pixel. 
\begin{figure}[t]
	\centering	
    \includegraphics[height=3.7cm]{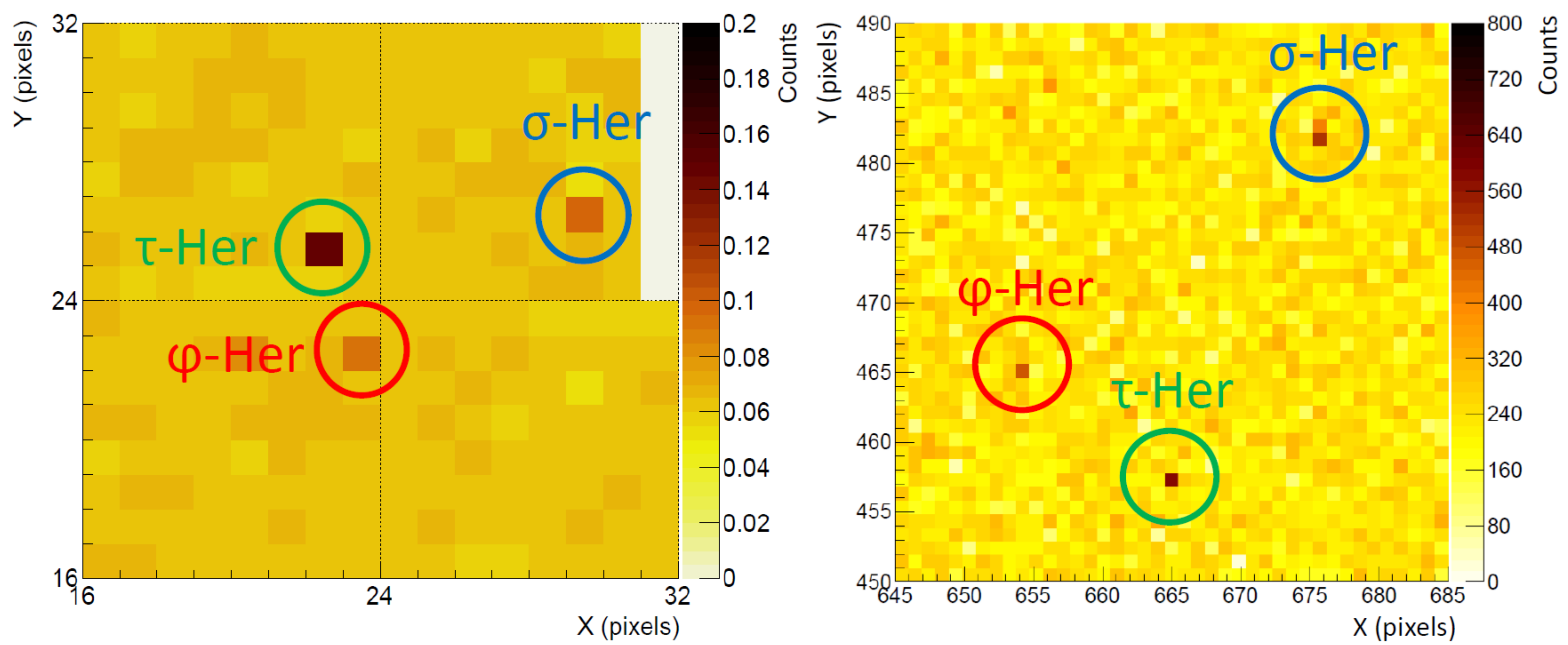}\vspace{2mm}
	\includegraphics[height=3.8cm, width=6cm]{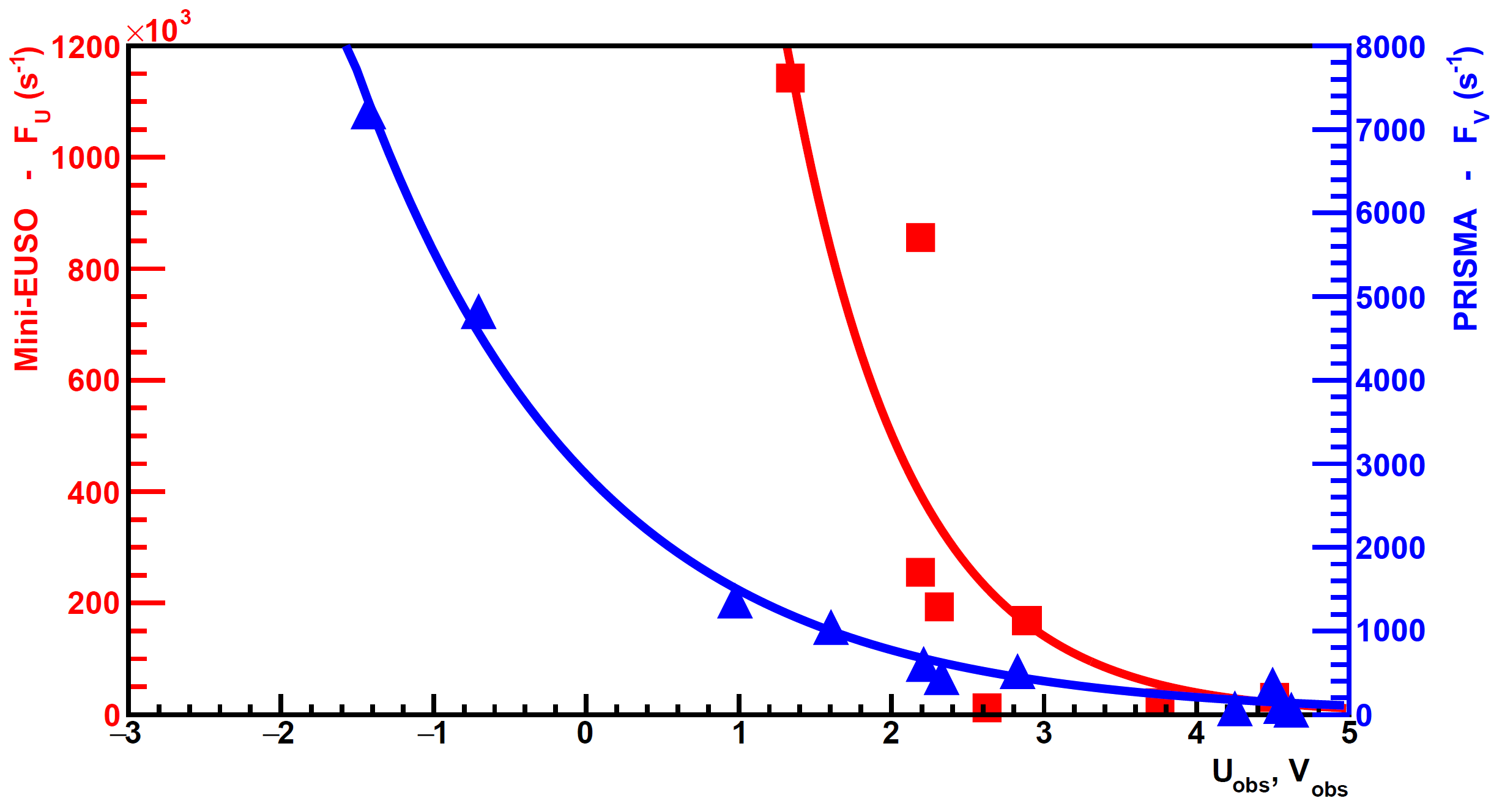}
	\caption{Three stars of the Hercules constellation. \textit{Left:} Stars in a Mini-EUSO~\textit{EM} data frame, integrated over 40.96~ms; \textit{Middle:} Same stars in an image of the PRISMA camera, integrated over 5~s. \textit{Right:} Flux of counts from stars and Jupiter for Mini-EUSO~\textit{EM} (left y-axis) and for the PRISMA camera (right  y-axis), vs. the U (for Mini-EUSO~\textit{EM}) and V (for PRISMA) apparent magnitudes.} \label{fig:stars}
\end{figure}
Figure~\ref{fig:stars} shows in the left panel the data frame of the Mini-EUSO~\textit{EM} with the Hercules stars, integrated over 40.96~ms. The same stars detected by the PRISMA camera, for which the data are integrated over 5~s, are shown in the middle panel. The PRISMA camera observes the sky with a spatial resolution of $\sim$\ang{0.2}/pixel, with a large wavelength bandwidth peaked in the visual. The right panel shows the flux of counts from the observed stars and Jupiter for Mini-EUSO~\textit{EM} (left red y-axis) and for the PRISMA camera (right blue y-axis), vs the U (for Mini-EUSO~\textit{EM}) and the V (for PRISMA) apparent magnitudes. The apparent magnitudes have been corrected for the atmospheric attenuation. Data points have been fitted with a exponential function with base 10, as the magnitudes are proportional to the logarithm with base 10 of the flux. In general, Mini-EUSO~\textit{EM} resulted to be sensitive to apparent magnitudes up to $\sim$4. The observation of the same stars from the same place and at the same time, i.e. with the same atmospheric condition, allows to compare the observations made with the two detectors. This would be useful in the eventuality of simultaneous observation of meteors with the PRISMA cameras and Mini-EUSO from space. Indeed, usually meteors are observed in the visual band, and the observation in the UV band could give more information about the development of meteors in the atmosphere, as well as their detection from both space and ground. 
\subsubsection{Meteors}
Four possible meteor events have been detected while Mini-EUSO~\textit{EM} was pointing to zenith at the Astronomical Observatory. From the comparison with the brightness of stars, they have apparent magnitudes of about 4. There is no counterpart of these events in the data of the PRISMA camera, as it is not sensitive to such faint meteors. However, as the meteor candidates developed at zenith, it is possible to estimate the horizontal component of the speed, supposing the development of meteors starts at $\sim$100~km altitude. The horizontal component of the speed resulted to be of the order of tens~km/s, consistent with the actual meteor development speed. 
\begin{figure}[h]
	\centering	
	\includegraphics[height=4.cm]{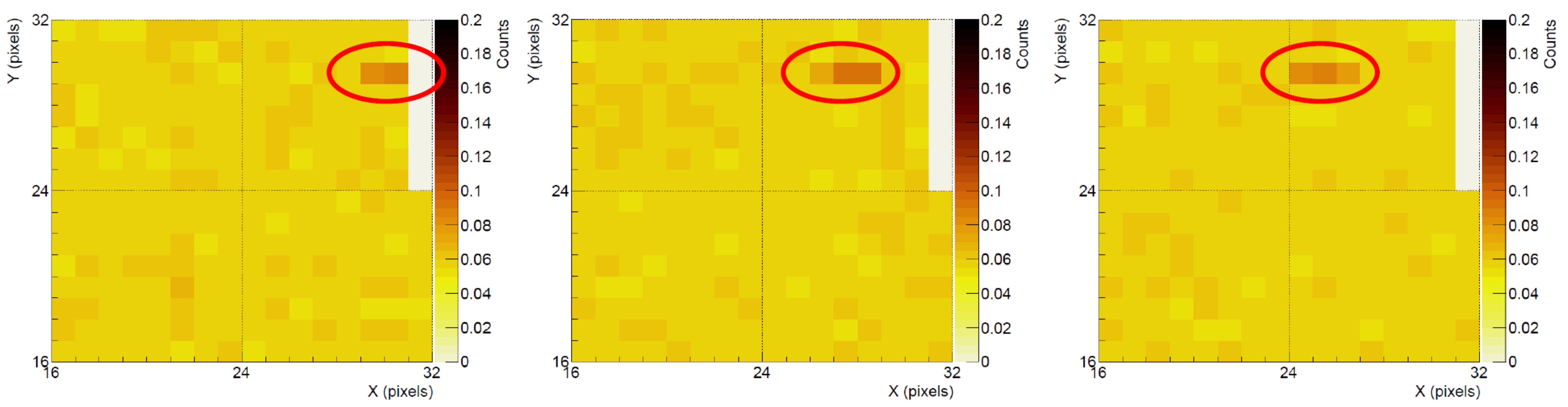}
	\caption{Meteor event in three data frames integrated over 40.96~ms.} \label{fig:meteor}
\end{figure}
Figure~\ref{fig:meteor} shows one of the candidate meteors, moving through three data frames integrated over 40.96~ms in the top-right MAPMT, from right to left.
\newline
\subsubsection{Background}
The observation of the sky from the Astronomical Observatory lasted up to sunrise. The increase of the luminosity is clearly detected by both Mini-EUSO~\textit{EM} and the PRISMA camera.
\begin{figure}[h]
	\centering
	\includegraphics[height=5cm]{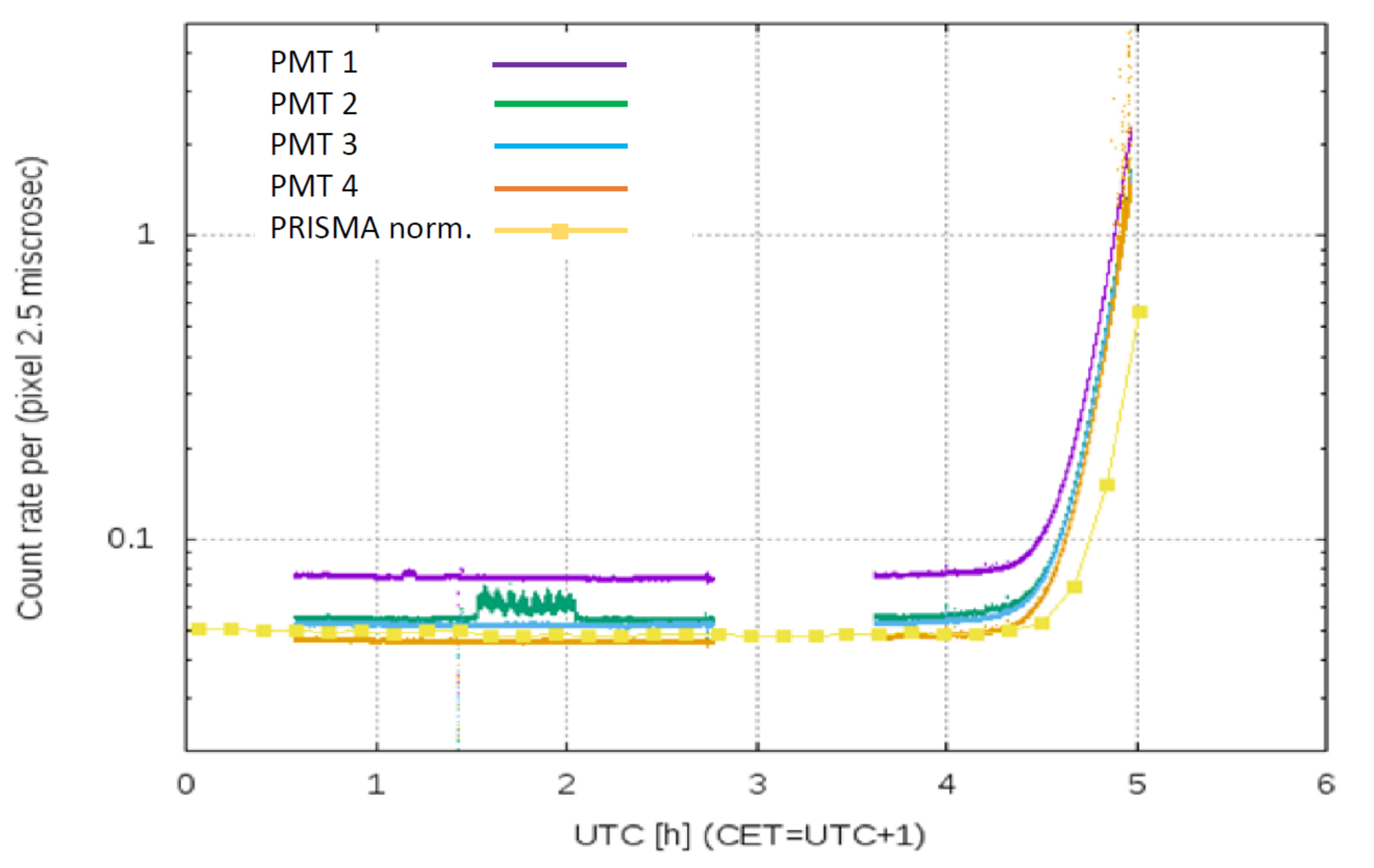}
	\caption{Background count rate over time of the four MAPMTs of Mini-EUSO~\textit{EM} and of the PRISMA camera, normalized to the mean value of the three MAPMTs with lower counts.} \label{fig:background}
\end{figure}
Figure~\ref{fig:background} shows the background count rate of the four MAPMTs of Mini-EUSO~\textit{EM} and of the PRISMA camera over time. The latter is normalized to the mean value of the three MAPMTs with lower counts. At sunrise, the counts increased first in UV and then in visual band, as UV photons are more scattered than visible ones, so stronger scattered light would be observed in UV. Peaks in the green line between 01:30 and 02:00~UTC correspond to a star crossing eight pixels of a MAPMT; the valleys correspond to the gap between adjacent pixels.

\subsection{Artificial sources}

\subsubsection{Lighted signs, flashes and city light}
Artificial lights from the urban area could be also used to test the detector. From the roof of the Department of Physics, lighted signs and flashes of skyscrapers and towers located in the Turin area (at $2-4$~km distance) were detected, as well as lights from Moncalieri and Chieri cities from the Astronomical Observatory. 
\begin{figure}[h]
	\centering
	\includegraphics[height=4.cm,width=3.cm]{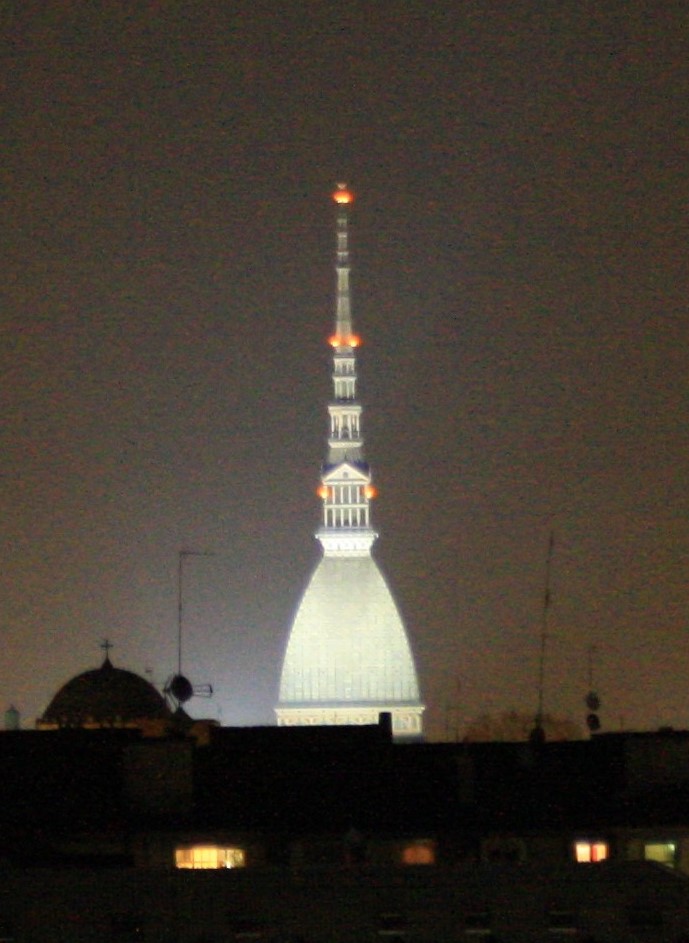}\hspace{0mm}
	\includegraphics[height=4.cm,width=3.cm]{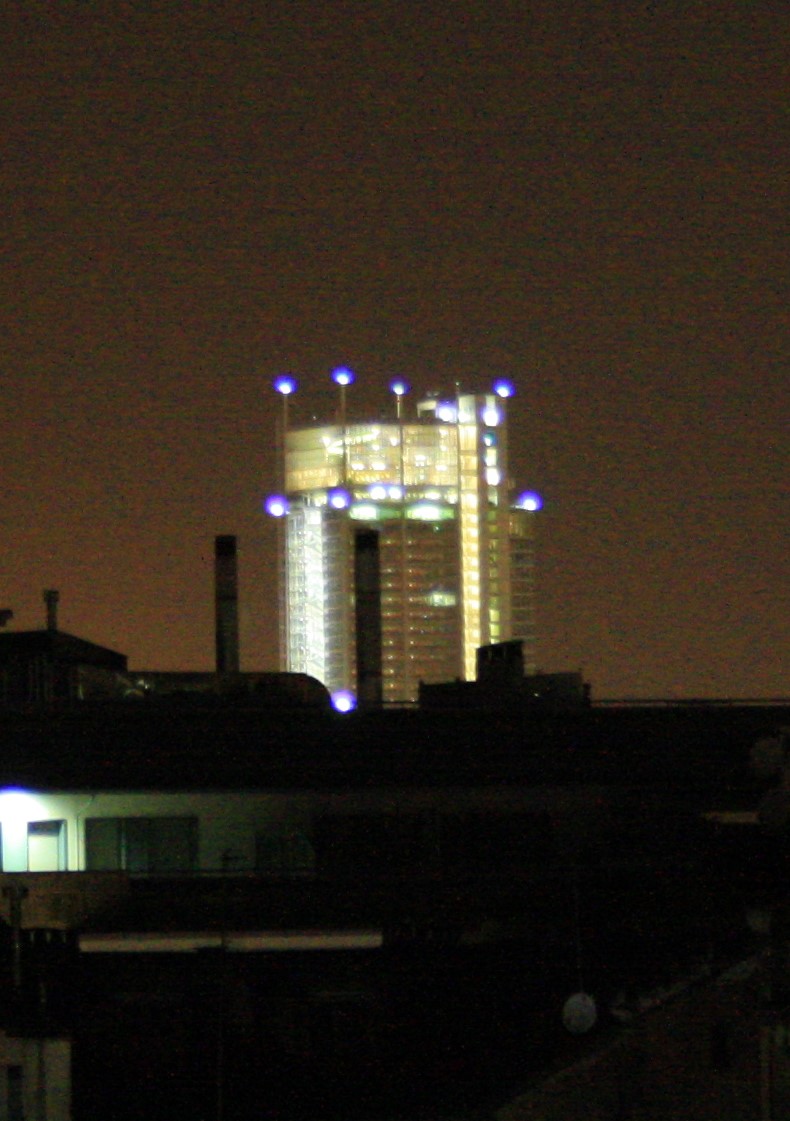}\hspace{5mm}	
	\includegraphics[height=4.cm]{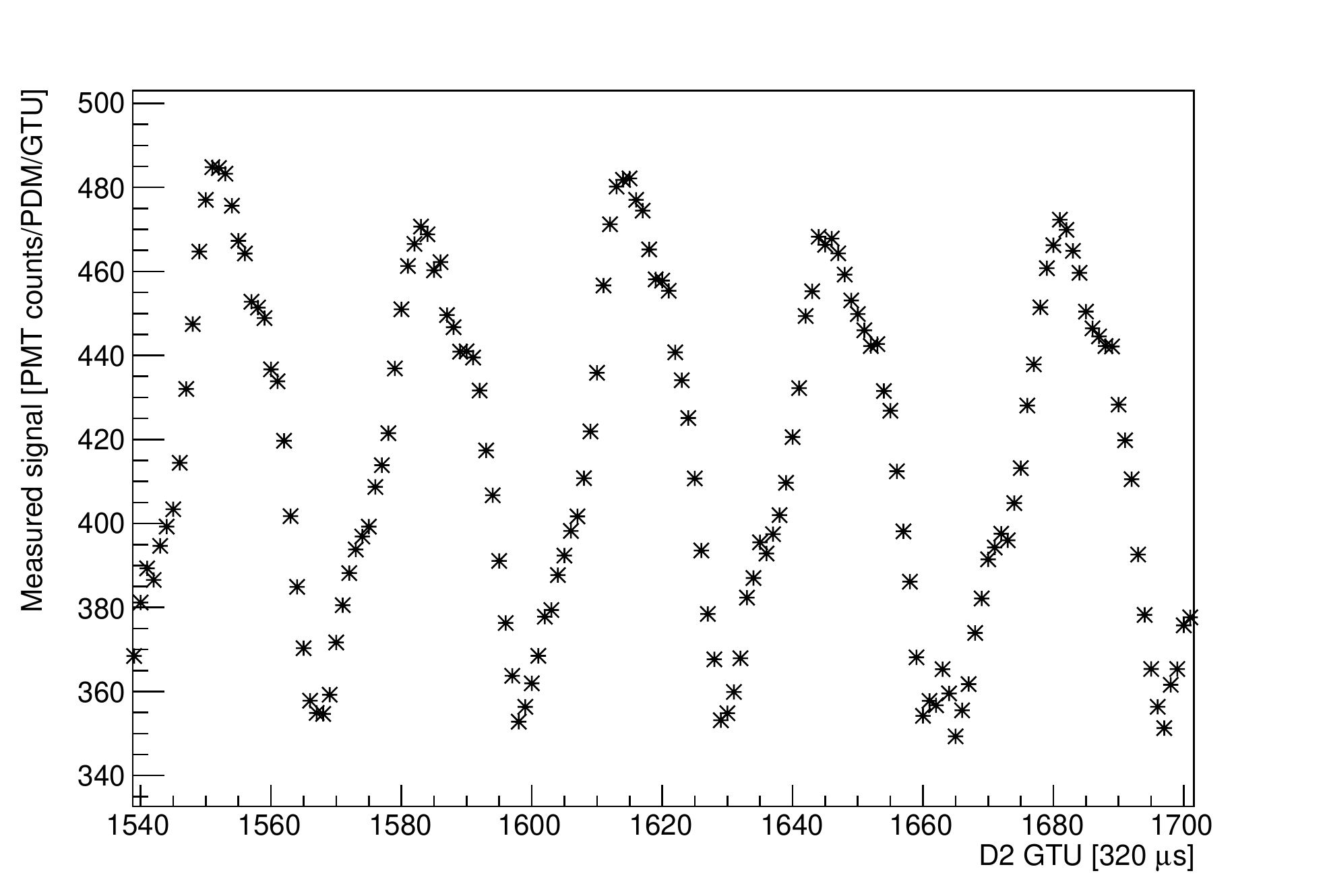}
	\includegraphics[height=4.cm]{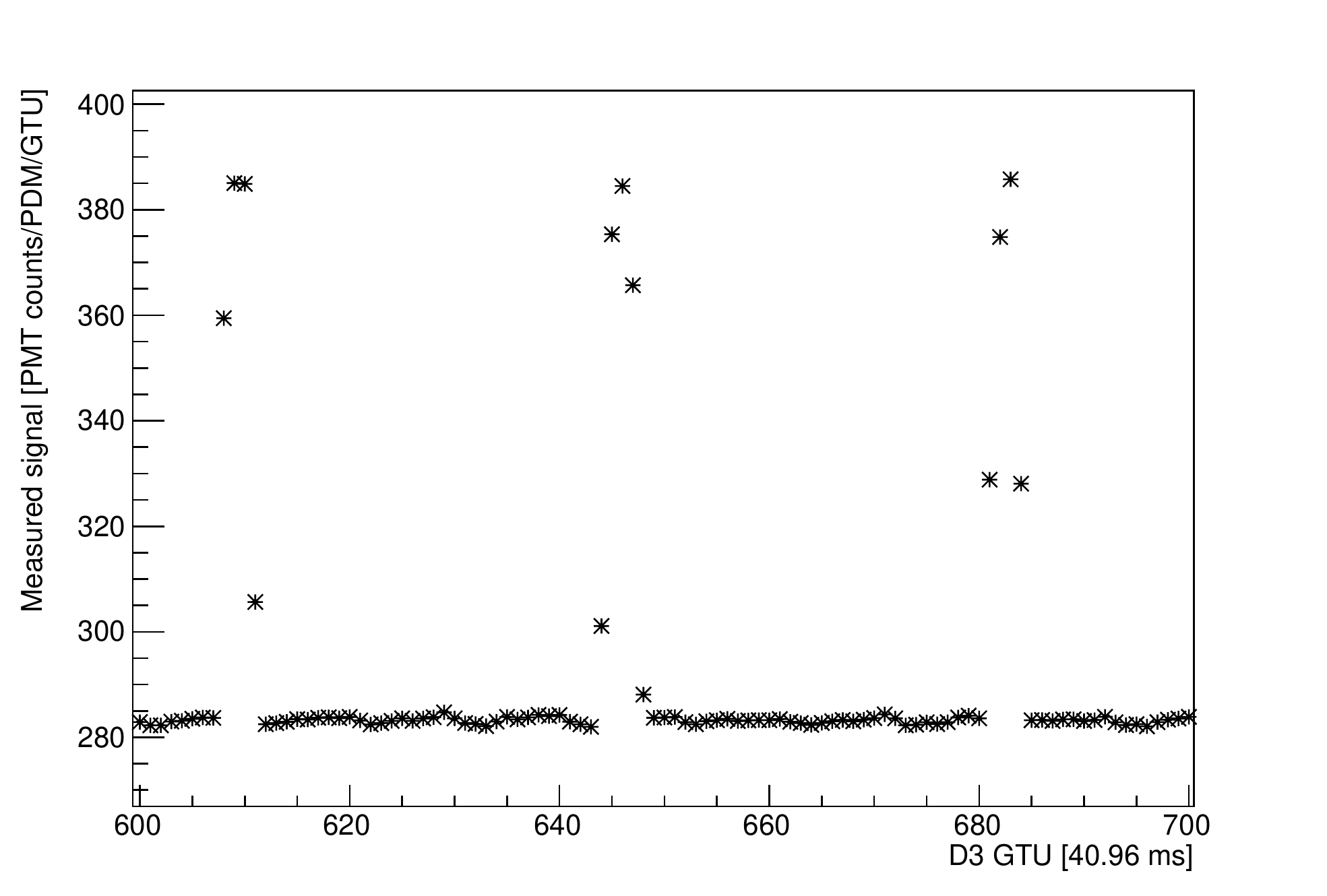}\hspace{6mm}	
	\includegraphics[height=4.cm]{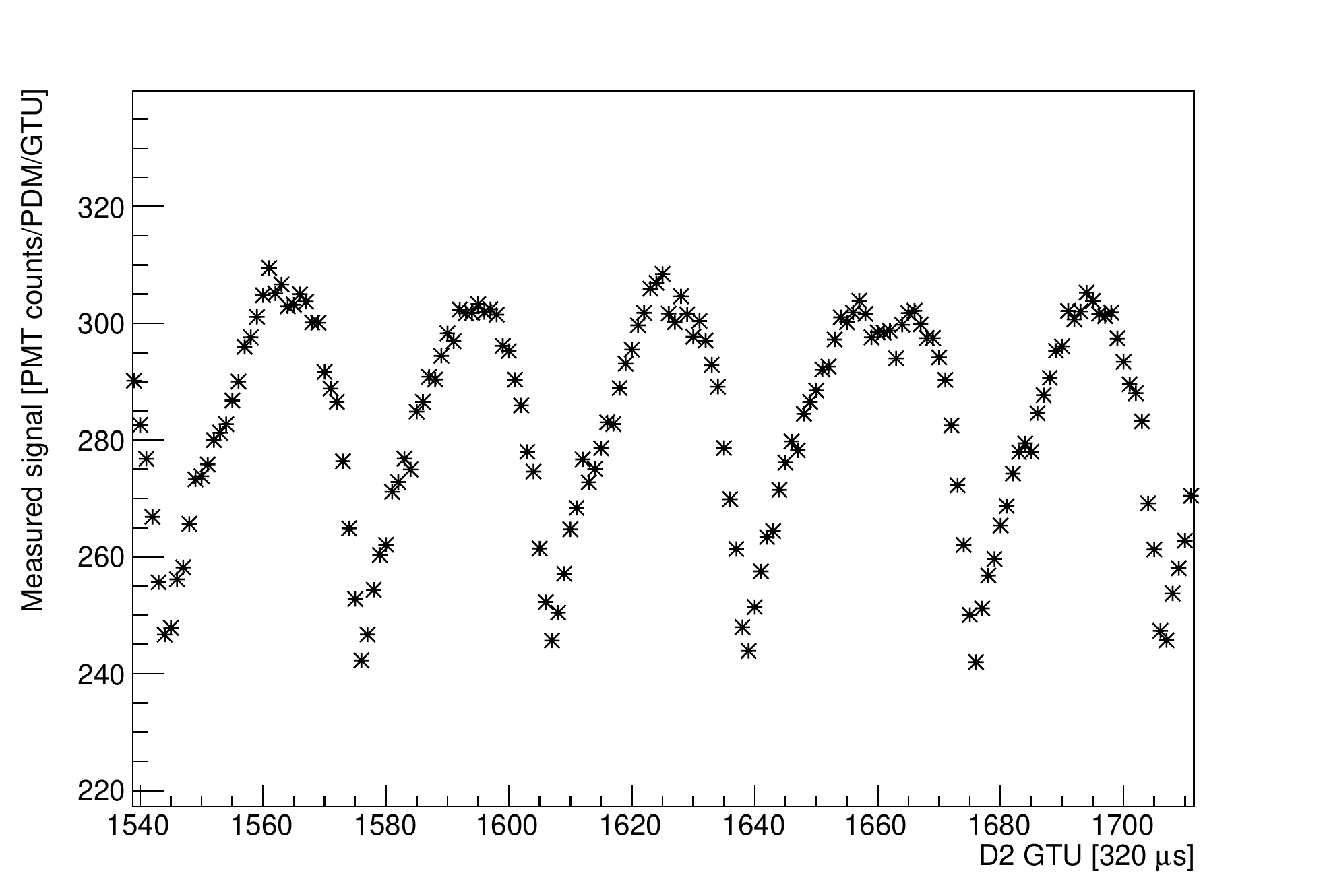}
	\caption{"Mole Antonelliana" tower and the "Intesa Sanpaolo" skyscraper. \textit{Top-left:} Pictures of the two buildings. \textit{Top-right:} Light-curve of the "Mole Antonelliana" tower with time bins of 320~$\mu$s showing the alternating signal with frequency $\sim$50~Hz. \textit{Bottom-left:} Light-curve with time bins of 40.96~ms showing flashes of the "Intesa Sanpaolo" skyscraper with frequency of $\sim$0.7~Hz; \textit{Right:} Light-curve with time bins of 320~$\mu$s showing the alternating signal with frequency $\sim$50~Hz.} \label{fig:buildings}
\end{figure}
Figure~\ref{fig:buildings} shows in the top-left panel pictures of the "Mole Antonelliana" tower and the "Intesa Sanpaolo" skyscraper. On the top-right, the light-curve of the "Mole Antonelliana" with time bins of 320~$\mu$s shows an alternating signal, with period $\sim$35~bins, corresponding to a frequency of $\sim$50~Hz, which is consistent with the frequency of the public electricity. At the bottom, the left panel shows the light-curve of the "Intesa Sanpaolo" skyscraper with time bins of 40.96~ms, where high pulses due to flashes are well above the baseline, pulsating with a frequency of $\sim$0.7~Hz. The right panel shows the light-curve with time bins of 320~$\mu$s, again with an alternating trend with frequency $\sim$50~Hz. This high frequency feature is present on the baseline and on the peaks as well. Other buildings with flashes have similar pulsating light-curves, and also generic light from cities have light-curves with such alternating profile. It will be important to see if the alternating signal will be present also from space, as it will indicate anthropogenic sources of background from the Earth. 

\subsubsection{Rocket of a telecommunication satellite and airplane}
\begin{figure}[ph]
	\centering	
	\includegraphics[height=4.cm]{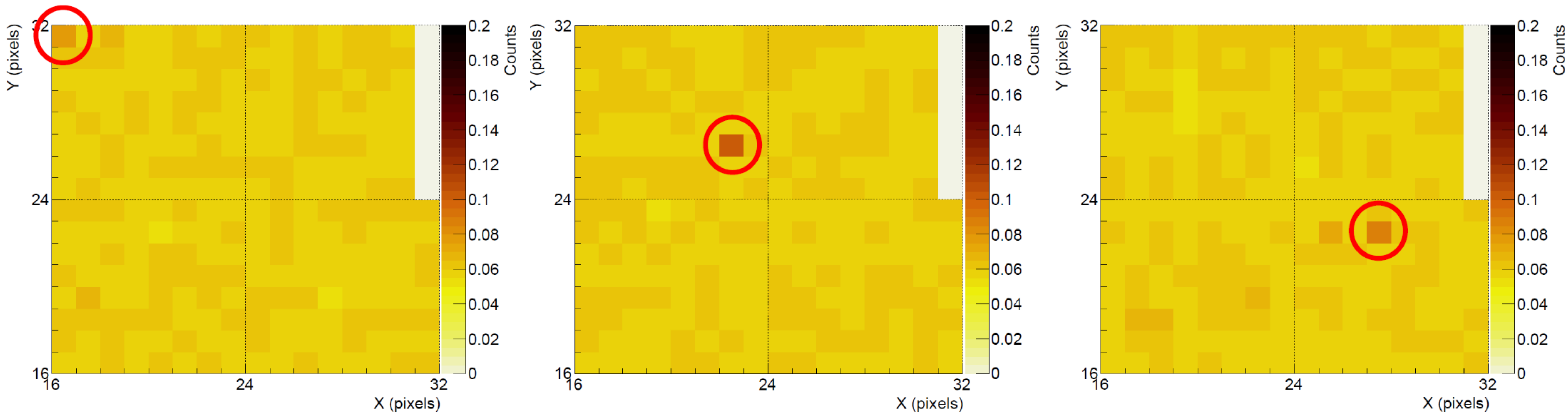}
	\caption{The rocket "Meteor 1-31 Rocket" in frames integrated over 40.96~ms.} \label{fig:satellite}
\end{figure}

During the observations at the Astronomical Observatory, while pointing to zenith, the orbiting rocket body that transported a telecommunication satellite was detected on March 14, 2018 at 04:10:58 UTC and later identified as the "Meteor 1-31 Rocket". It is a 2.6~m$\times$2.8~m large-size debris orbiting at an average height of $\sim$550~km and an angular velocity of \ang{0.78}/s. Figure~\ref{fig:satellite} shows the satellite as a bright pixel moving from the top-left corner towards the center of the frame in three data frames integrated over 40.96~ms. Size and distance of this satellite have been rescaled in order to reproduce the detection of space debris with Mini-EUSO on-board the ISS. Also simulation studies have been done for this purpose \cite{bib:debris}. This study is important to verify the working principle of a space-based system under development for the detection and remediation of space debris \cite{bib:debris_detector}. 

\begin{figure}[ph]
	\centering	
	\includegraphics[height=5.cm]{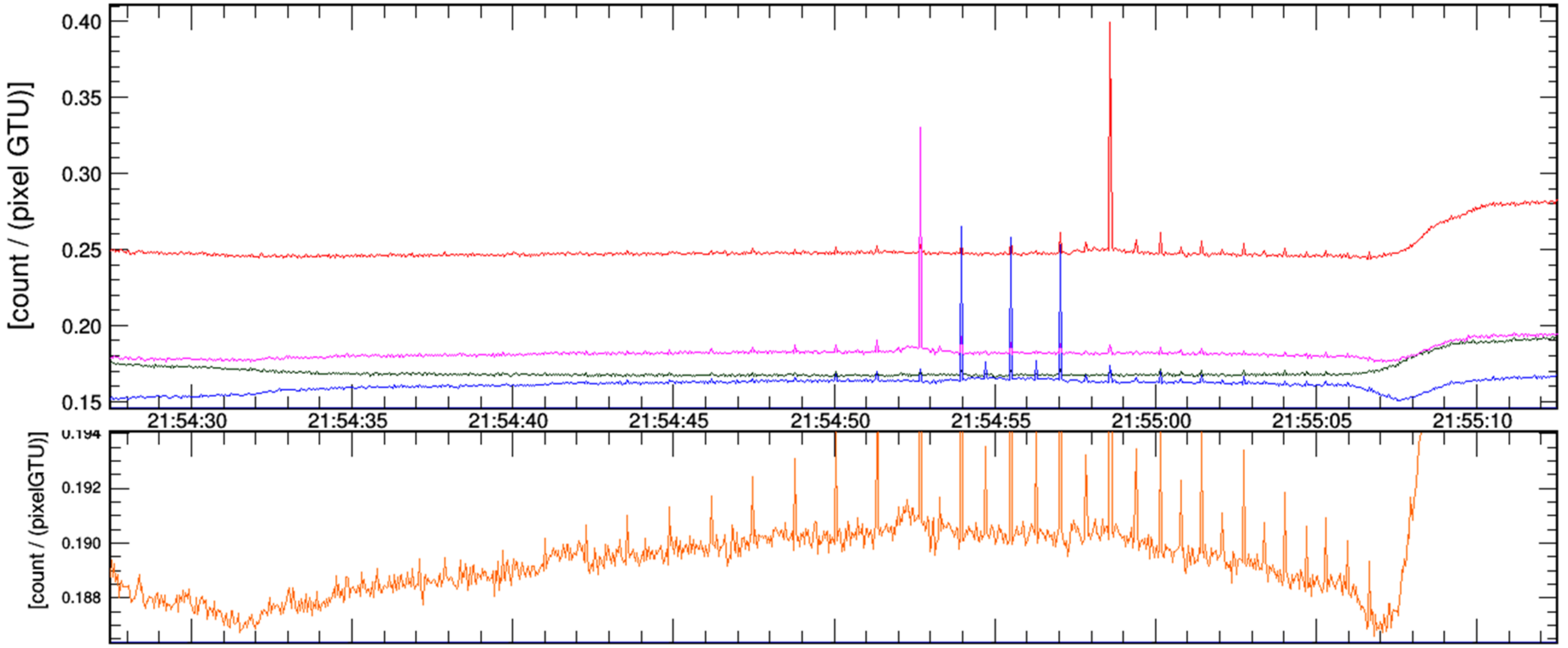}
	\caption{Airplane for the flight LH1902 detected on March 12 2018 in the time interval 21:54:27-21:55:12 UTC. \textit{Top:} Peaks from four separate MAPMTs; \textit{Bottom:} Peaks from the whole EC.} \label{fig:airplane}
\end{figure}

Also flashes of the airplane performing the flight LH1902 from Munich to Turin have been detected, from the same site, on March 12 2018 in the time interval 21:54:27-21:55:12 UTC (about 6 minutes before landing). It passed at an altitude $\sim$780-690~m above the observation site at 610~m a.s.l., with a speed of 426~km/h. Figure~\ref{fig:airplane} shows peaks from four separate MAPMTs (top panel) and from the whole EC (bottom panel).
\section{Summary and Conclusion}
The observations made with the Mini-EUSO~\textit{EM} offered the possibility to test successfully the prototype of the Mini-EUSO detector that will be launched and installed on the ISS this year, 2019. They provided a large variety of data on different timescales, to test the detector performance. Stars with apparent magnitude up to $\sim$4 allowed to make a comparison of the performance of Mini-EUSO~\textit{EM} and the PRISMA camera of the Astronomical Observatory of Turin. The possible detection of meteors showed that the telescope is able to detect faint meteors with speeds of $\sim$tens km/s. The analysis of the background around sunrise time showed that the UV photons are more scattered than the visual ones. Observations of artificial light from buildings showed the capability of detecting signals with frequencies of 0.7~Hz and 50~Hz. Finally, with the detection of a space rocket it was possible to rescale the size and the distance to the detector to emulate the detection of space debris with Mini-EUSO from space. 

\newpage

\section*{Acknowledgments}
\noindent The support received by the Astronomical Observatory of Turin is deeply acknowledged.\\
This work was partially supported by Basic Science Interdisciplinary Research Projects of RIKEN and JSPS KAKENHI Grant (JP17H02905, JP16H02426 and JP16H16737), by the Italian Ministry of Foreign Affairs and International Cooperation, by the Italian Space Agency through the ASI INFN agreement n. 2017-8-H.0 and contract n. 2016-1-U.0, by NASA award 11-APRA-0058 in the USA, by the Deutsches Zentrum f\"ur Luft- und Raumfahrt, by the French space agency CNES, the Helmholtz Alliance for Astroparticle Physics funded by the
Initiative and Networking Fund of the Helmholtz Association (Germany), by Slovak Academy of Sciences MVTS JEM-EUSO as well as VEGA grant agency project 2/0132/17, by National Science Centre in Poland grant (2015/19/N/ST9/03708 and 2017/27/B/ST9/02162), by Mexican funding agencies PAPIIT-UNAM, CONACyT and
the Mexican Space Agency (AEM). Russia is supported by ROSCOSMOS and the Russian Foundation for Basic Research Grant n. 16-29-13065. Sweden is funded by the Olle Engkvist Byggm\"astare Foundation.


\end{document}